\documentclass[twocolumn,showpacs,preprintnumbers,amsmath,amssymb,aps,prl]{revtex4-1}

\usepackage{graphicx}

\usepackage{dcolumn}% Align table columns on decimal point

\usepackage{bm}% bold math

\usepackage{color}

\begin{document}

\title{Beyond Anderson Localization in 1D:  Anomalous Localization of Microwaves in  Random  Waveguides}

\author{A.  A. Fern\'andez-Mar\'in$^1$, J. A. M\'endez-Berm\'udez$^1$, J. Carbonell$^2$, F.  Cervera$^2$, J.   S\'anchez-Dehesa$^2$, and V.  A. Gopar$^3$}
\affiliation{$^1$Instituto  de F\'isica, Benem\'erita Universidad Aut\'onoma de Puebla, Apartado Postal J-48, Puebla 72570, Mexico}
\affiliation{$^2$Wave Phenomena Group, Universitat Polit\`ecnica de Val\`encia, Camino de vera s. n. (Edificio 7F), ES-46022, Valencia, Spain}
\affiliation{$^3$Departamento de F\'isica Te\'orica, Facultad de Ciencias, and Instituto de Biocomputaci\'on y F\'isica de Sistemas Complejos, Universidad de Zaragoza, Pedro Cerbuna 12, E-50009, Zaragoza, Spain.}

\begin{abstract}

Experimental evidence demonstrating that anomalous localization of waves can be induced in a controllable manner is reported. A microwave waveguide with dielectric slabs randomly placed is used to confirm  the presence of anomalous localization.  If the random spacing between slabs follows a distribution with a power-law tail (L\'evy-type distribution), unconventional properties in the  microwave-transmission fluctuations  take place revealing the presence of anomalous localization. We study both theoretically and experimentally the complete distribution of the transmission through random waveguides characterized by  $\alpha=1/2$ (``L\'evy waveguides'') and $\alpha=3/4$, $\alpha$ being the exponent of the power-law tail of the L\'evy-type distribution. As we show, the  transmission distributions are determined by only two parameters, both of them experimentally accessible. Effects of anomalous localization on the transmission are compared with those from the standard Anderson localization.

\end{abstract}

\pacs{41.20.Jb,42.25.Dd, 42.25.Bs,  05.40.-a}

\maketitle

Wave localization in random media occurs as a consequence of  coherent destructive interference  in multiple scattering and plays a central role in the description of transport of quantum and classical waves. The physical mechanism behind localization,  introduced by Anderson in quantum electron transport \cite{anderson}, is so general that the study of localization has not been restricted to electrons but light, sound waves, microwaves, and ultracold atoms have been considered in theoretical and  experimental investigations  \cite{today,50years}. 

The exponential spatial decay of the wave envelope, named Anderson localization, is the most studied type of localization and its effects can be recognized by means of transport quantities, such as the transmission. Hence, several properties of the transmission through disordered structures, like random waveguides and disordered photonic structures, have been experimentally studied in the presence of Anderson localization \cite{wiersma,stoytchev,chabanov,genack,penha,yamilov}. Electron-electron interactions or correlated disorder potentials, however, affect the observation of Anderson localization. Actually, it is known that correlated disorder leads to anomalous localization; i.e., correlations may enhance or reduce the wave localization, in relation to the Anderson localization. Extensive literature on the problem of localization with correlated disorder already exists \cite{izrailev}.

Therefore, it is  widely known that the presence of (uncorrelated) disorder leads to Anderson localization of waves in one-dimensional systems. It is less known, however, that waves can be localized in a weaker and different way  than in the standard Anderson localization,  even in the absence of correlations in the potentials. For instance, concerning  the problem of quantum electronic transport, within a single electron picture,  numerical simulations  have shown a power-law decay  of  the conductance versus the system length in  disordered 1D wires at the band center \cite{soukoulis, evangelou,elias}, which is in contrast to the faster exponential decay predicted for quantum wires in the presence of  Anderson localization. Also,  new advances  in fabrication of photonic structures have allowed the experimental observation  of anomalous localization in disordered glasses \cite{barthelemy}.

From a practical point of view, the degree of disorder   might be manipulated to engineer the properties of materials  \cite{salman,riboli}. In this sense, disordered optical fibers have been recently fabricated to transport high-quality images,  taking advantage of light localization \cite{salman}. Therefore, controlling  effects of disorder, such as  localization, is of practical importance.

Motivated by a recent proposal \cite{falceto_gopar} to induce  anomalous localization of electron wave functions  by considering  L\'evy-type disorder (defined below) in 1D quantum wires, this Letter demonstrates that anomalous localization of electromagnetic waves can be induced in a controllable manner in 1D waveguides. The presence of anomalous localization leads to unconventional properties of the transmission, as we show below. 
\begin{figure}
%\begin{center}
 \includegraphics[width=1.2\columnwidth]{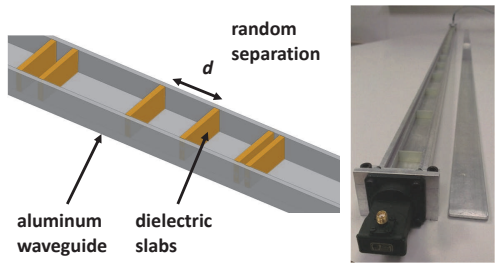}
 \caption{(Color online). Schematic of a random waveguide  with scatterers (dielectric slabs) randomly placed according  
 to a L\'evy-type distribution (left). Experimental setup employing an $X$-band waveguide (top plate open to allow inner vision). Dielectric  slabs act as scattering elements of the transmitted signal (right).}
 \label{figure_1}
% \end{center}
\end{figure}

The  experiments are performed in the microwave regime using an array of dielectric slabs whose nearest-neighbor separation $d$ follows a probability density function with a power-law tail.  As described below, the observed averages of the microwave transmission depend  on the exponent $\alpha$  of the power-law tail through the relationships $\langle T \rangle  \propto   L^{-\alpha}$ and $\langle -\ln T \rangle \propto  L^\alpha$, $L$ being the length of the waveguide. Thus, $\alpha$ can be used to characterize the strength of the anomalous localization. The complete transmission distribution is experimentally obtained from different disorder realizations for the cases $\alpha=1/2$ and 3/4,  at two different microwave frequencies. Transmission distributions in the presence of  Anderson localization are also experimentally obtained for the sake of comparison.

{\it Theoretical model}.-  We briefly introduce the main ideas and results of the  model in  Ref.~\cite{falceto_gopar}. First, we  assume  that waves can travel coherently through a disordered  waveguide, where the source of disorder  is introduced  by placing scatterers randomly,  and independently, separated, as we show schematically in Fig.~\ref{figure_1}. We also assume  that waves  propagate in one dimension \cite{models}. The special feature of the system  we are considering is that the probability density of spacing $\rho(d)$ between scatterers has a power-law tail;  i.e.,  for large $d$, $\rho(d) \sim c/d^{1+\alpha}$, where $c$ is a constant and $0<\alpha<1$ \cite{largervalues}. We notice that because of  the power-law tail,  the first and second moments of $\rho(d)$ diverge. Those heavy-tailed probability densities, named $\alpha$-stable distributions (also referred to as L\'evy-type distributions), have been objects of mathematical interest for a long time \cite{levy,kolmogorov,uchaikin}.

Assuming a  spacing probability density  with a heavy tail like $\rho(d)$,   the probability density $Q_L(n)$ of the number of scatterers $n$ in a system of  fixed length  $L$  is given by \cite{falceto_gopar}
\begin{equation}\label{Qalpha}
Q_{L}(n)=\frac{2L}{\alpha} {(2 n)^{-{(1+\alpha)}/{\alpha}}} q_{\alpha,c}[{L}/{(2 n)^{1/\alpha}}] ,
\end{equation}
where the probability density function   $q_{\alpha,c}$ is a  L\'evy-type  distribution characterized by a power-law tail with exponent $\alpha$, like $\rho(d)$  for large values of $d$. Having the probability density $Q_L(n)$, we can calculate the distribution of the transmission of  our ``anomalous  waveguides'' by using the following  result from the standard scaling approach to localization \cite{melnikov, abrikosov, dorokhov,mello_groups}: For  $n$ average number of scatterers, the transmission probability  density $p_s(T)$ in 1D  is given by     
\begin{equation}\label{pofg}
p_s(T)=\frac{s^{-{3}/{2}}}{\sqrt{2\pi}}
\frac{e^{-{s}/{4}}}{T^2}\int_{y_0}^{\infty}dy\frac{y{\rm e}^{-{y^2}/{4s}}}
{\sqrt{\cosh{y}+1-2/T}},
\end{equation}
where $y_0={\rm arcosh}{(2/T-1)}$ and $s=an$,  $a$ being a constant. The parameter $s$ can also be  identified as the length of the system in units of the mean free path. Thus, combining  Eqs.~(\ref{Qalpha}) and (\ref{pofg}),  we obtain the transmission distribution $P(T)$ for random waveguides of length $L$, whose probability density of spacing  between scatterers has a power-law tail: $P(T)=\int_{0}^\infty p_{s}(T)Q_{L}(n){d}n $. Substituting $Q_{L}(n)$, Eq.~(\ref{Qalpha}), in the previous integral expression for $P(T)$  and using the scaling properties of  L\'evy-type distributions, we  finally write  $P(T)$ as 
\begin{eqnarray}\label{pofT_xi}
P(T)=\int_0^\infty p_{s(\alpha,\xi,z)}(T) q_{\alpha,1}(z){\rm d}z ,
\end{eqnarray}
where we have defined the variables $z=L/(2n)^{1/\alpha}$ and  $s(\alpha,\xi,z)={\xi}/(2{z^\alpha I_\alpha)}$, with  $\xi=\langle -\ln T \rangle$ and $I_\alpha=(1/2)\int_{0}^{\infty}z^{-\alpha} q_{\alpha,1}(z) dz$. We remark that the distribution $P(T)$ [Eq.~(\ref{pofT_xi})] only depends on two parameters: $\alpha$ and $\xi$, which means that all other details of the disorder configuration are irrelevant.

From the transmission distribution, Eq.~(\ref{pofT_xi}), one can obtain the previously mentioned  power-law behavior of  the ensemble average of the transmission,
\begin{equation}\label{averageT}
\langle T \rangle \propto L^{-\alpha},
\end{equation}
and  the average of the logarithm of the  transmission,  
\begin{equation}
\label{averagelnT}
\langle - \ln T \rangle \propto L^\alpha, 
\end{equation}
for $0<\alpha <1$. We point out that Eqs.~(\ref{averageT}) and (\ref{averagelnT}) are in contrast to the known results in the Anderson localization problem: $\langle T \rangle$ decays exponentially with $L$  and $\langle -\ln T \rangle \propto L$. Therefore,  the above  nonlinear dependencies  of $\langle T \rangle $ and $ \langle -\ln T \rangle$  on $L$ reveal  the presence of anomalous localization.

In summary, according to the above described model, we can induce anomalous localization in a waveguide by randomly placing scatterers whose  separations follow a probability density  with a power-law tail. Indeed, the exponent $\alpha$ of the power-law tail determines the  strength of the anomalous localization. If additionally the information of the value of the ensemble average $\langle -\ln T \rangle$ is known, we can obtain   the complete distribution of the  transmission from  Eq.~(\ref{pofT_xi}).

Numerical simulations were performed firstly in order to  design  the random waveguides and give additional support to the experimental results. The transmission through the waveguide is calculated by using a transfer matrix method \cite{markos}. We consider two cases for the probability density  of spacing $\rho(d)$ between slabs: $\alpha=1/2$ and $3/4$. In particular, for $\alpha=1/2$, we use the so-called L\'evy distribution given by $\rho(d)=q_{1/2,1}(d)=(1/\sqrt{2\pi})d^{-3/2}\exp{(-1/2d})$. For $\alpha=3/4$, there is no analytical expression for $q_{3/4,1}(d)$, but it can be numerically computed \cite{uchaikin}.

\begin{figure}
\includegraphics[width=0.95\columnwidth]{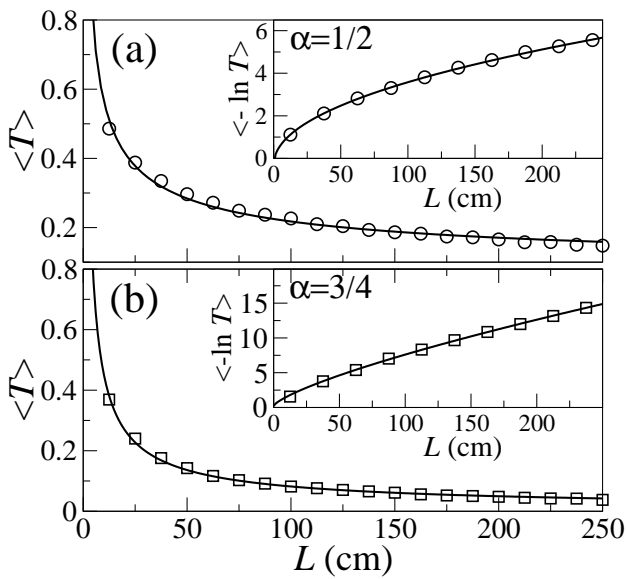}
\caption{Numerical simulation results (symbols) for the ensemble averages $\langle T \rangle$ (main panel) and $\langle -\ln T \rangle$ (insets) as a function of the length $L$ of the waveguides. Fits (solid lines) to the numerical data are obtained according to Eqs.~(\ref{averageT}) and (\ref{averagelnT}). In agreement with the theoretical model, (a) $\langle T \rangle \propto L^{-1/2}$ and $\langle -\ln T \rangle \propto L^{1/2}$ for L\'evy waveguides, while (b) $\langle T \rangle \propto L^{-3/4}$ and $\langle -\ln T \rangle \propto L^{3/4}$ for $\alpha=3/4$.} 
\label{figure_2}
\end{figure}

Thus, we  generate numerically an ensemble of random waveguides and obtain the averages $\langle T \rangle$ and  $\langle -\ln T \rangle$ over 40~000 disorder realizations at different values of the length $L$ of the waveguides at 8.5 GHz. The  symbols in Figs.~\ref{figure_2}(a) and 2(b)  (main panels) show the power-law behavior of $\langle T \rangle$ with $L$ for random waveguides with spacing densities characterized by $\alpha=1/2$ and $\alpha=3/4$, respectively. As predicted by the model, Eq.~(\ref{averageT}), for  $\alpha=1/2$,  we found that $\langle T \rangle$ depends on $L$ as $L^{-1/2}$ [solid line in Fig.~\ref{figure_2}(a)], while for  $\alpha=3/4$, $\langle T \rangle$ behaves as $L^{-3/4}$ [solid line in Fig.~\ref{figure_2}(b)]. Similarly, the insets in Fig.~\ref{figure_2} show the numerical results (symbols) for the ensemble average $\langle -\ln T \rangle$. In this case, the solid lines show that $\langle -\ln T \rangle \propto L^{1/2}$ [Fig.~\ref{figure_2}(a)]   while  $\langle -\ln T \rangle \propto L^{3/4}$  [Fig.~\ref{figure_2}(b)] for $\alpha=1/2$ and $\alpha=3/4$, respectively, which is in agreement with our previous result given by  Eq.~(\ref{averagelnT}).

With the above numerical support, we proceed with the experimental analysis of the  transmission distribution. 

{\it Experimental results and discussions}.- We consider a 2-m-long aluminum waveguide with randomly (and independently) placed dielectric slabs, as we show in Fig.~\ref{figure_1}. According to the previous model, the spacing between slabs is obtained by sampling from a L\'evy-type distribution with parameter $\alpha$. As in the above numerical simulations, we consider two values: $\alpha =1/2$ and 3/4. The case $\alpha=1/2$ corresponds to the so-called L\'evy distribution, so we name those  random waveguides ``L\'evy waveguides.'' Note that $\alpha$ is employed here to control experimentally the strength of the anomalous localization.

A standard microwave $X$-band transition (coaxial to waveguide, see Fig.~\ref{figure_1}) feeds the random waveguide and, on the opposite side, a second transition captures the transmitted signal. This makes it possible to obtain transmission and reflection coefficients from a vector network analyzer. We consider two excitation frequencies, 8.5 and 11.5 GHz. In both cases, we are in the frequency band where only the fundamental TE$_{10}$ mode propagates through the waveguide. To obtain the transmission statistics, we construct an ensemble of random waveguides of different disorder realizations and collect the transmission data across the ensemble. Experimentally, however,  it is a time-consuming task to perform a large number of different random configurations of the dielectric slabs. Therefore, in order to increase the size of the data collection, we consider transmission measurements in small frequency windows around 8.5 and 11.5 GHz. The size of these frequency windows (0.4 GHz, in both cases) is small enough that we can assume   that the statistical properties of the transmission do not change within  that frequency interval \cite{average}. Thus, we have performed 135 different disorder configurations. For each disorder realization we have measured the transmission at five frequency points around the two nominal frequencies (8.5 and 11.5 GHz). These data form a collection of 675  measurements at each nominal frequency.

\begin{figure}
\includegraphics[width=0.95\columnwidth]{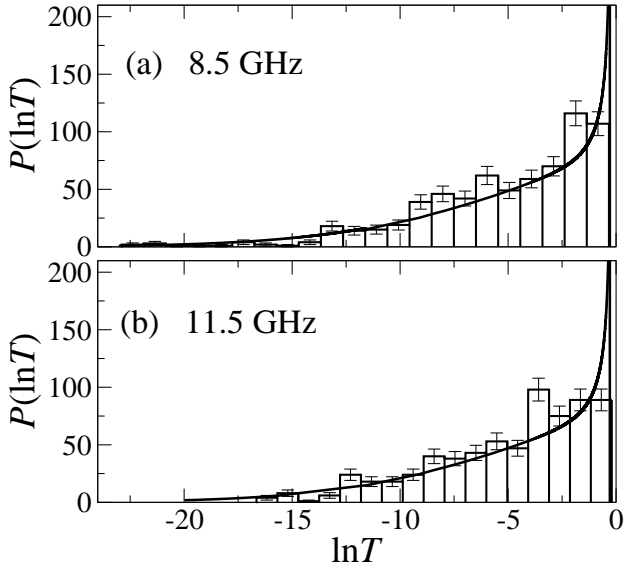}
\caption{Experimental distributions $P(\ln T)$ (histograms with finite-sample error bars) for L\'evy waveguides ($\alpha=1/2$) at two different nominal frequencies. The solid lines in (a) and (b) are obtained from Eq.~(\ref{pofT_xi}) with $\langle -\ln T \rangle =5.1$ and 4.9, respectively. Note that the distributions are not normalized to unity: the integral over $T$ gives the size of the sample (675). We can observe that the trend of the experimental distributions is well described by the theoretical  model.}
\label{figure_3}
\end{figure}

First, we consider the case of L\'evy waveguides ($\alpha =1/2$). In Figs.~\ref{figure_3}(a) and \ref{figure_3}(b) we show the experimental transmission distributions (histograms) at the nominal frequencies $8.5$ and $11.5$ GHz, respectively. Notice that since transmission through the waveguides is small, the distribution of the logarithm of the transmission $P(\ln T)$, instead of $P(T)$, is more meaningful in order to appreciate the details of the distributions. From the transmission data collection, we found  that $\langle -\ln T \rangle =5.1$ at  $8.5$ GHz, while $\langle - \ln T \rangle =4.9$ at  $11.5$ GHz. As mentioned previously, with the information of $\langle -\ln T \rangle$ and $\alpha$, we can obtain the theoretical distributions from Eq.~(\ref{pofT_xi}) or $P(\ln T)$  after the change of variable $T \to \ln T$. As we can observe from Fig.~\ref{figure_3}, both sets of experimental data are well described by the model (solid lines). The largest differences between theoretical and experimental distributions are seen near the origin, i.e.,  at $\ln T \approx 0$. Our numerical simulations indicate that those discrepancies are due to absorption losses limiting the maximum transmission \cite{losses}. Notice that the distributions show a small gap near the origin: this is due to absorption, which prevents perfect transmission $T=1$, or $\ln T =0$. We remark, however, that there are no free parameters in our theoretical  approach.

\begin{figure}
\includegraphics[width=0.95\columnwidth]{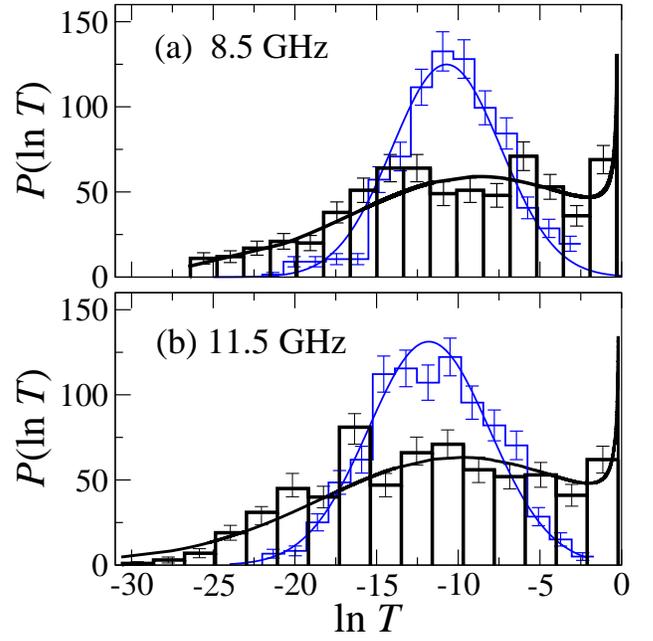}
\caption{(Color online). Experimental distributions $P(\ln T)$ (thick black solid line histograms with finite-sample error bars) for waveguides with random separations of slabs following  a density probability with $\alpha=3/4$, at two different nominal frequencies. The thick black solid curves  show the results from  Eq.~(\ref{pofT_xi}). Note that the distributions are not normalized to unity, as in Fig. \ref{figure_3}. A good agreement is seen between experiment and theory. Histograms in thin blue (gray)  solid line are experimental results for  random waveguides in the presence of standard Anderson localization, while the curves [thin blue (gray) solid lines] correspond to the expected log-normal distributions. Strong differences between the distributions of $\ln T$ for  anomalous and Anderson localizations are  seen.}
\label{figure_4}
\end{figure}

Now, we change the properties of the disorder by varying the parameter $\alpha$ of the spacing distribution of the dielectric slabs to $\alpha= 3/4$; i.e., we construct an ensemble of waveguides with a spacing distribution  that follows the  L\'evy-type distribution $\rho(d)= q_{3/4,1}(d)$. As in the previous case of $\alpha=1/2$, we follow  the same data collection procedure at the nominal frequencies $8.5$ and $11.5$ GHz. In Fig.~\ref{figure_4} we show the experimental transmission distribution $P(\ln T)$ (thick black solid line histograms). The experimental average values are $\langle -\ln T \rangle=10.6$ and $\langle - \ln T \rangle=11.8$ at the nominal frequencies $8.5$ and $11.5$ GHz, respectively. The black solid curves in Figs.~\ref{figure_4}(a) and \ref{figure_4}(b) are obtained from Eq.~(\ref{pofT_xi}) with $\alpha=3/4$ and using  the corresponding experimental values of $\langle -\ln T \rangle$. Once again, the model gives a good description of the experimental results.

It is  interesting to compare  the effects  of anomalous and  Anderson localizations on the transmission. We thus have constructed an ensemble of random waveguides with disorder that leads to Anderson localization. In particular, we have considered a spacing distribution of slabs that follows a Gaussian distribution. The parameters of the Gaussian distribution are chosen in such a way that the average values $\langle -\ln T\rangle$ are approximately the same as those obtained in the previous case of $\alpha =3/4$. The experimental results (thin gray solid  line histograms) are shown in Fig.~\ref{figure_4}. The thin line curves correspond to the known log-normal distribution expected for 1D systems in the presence of   Anderson localization. Strong differences are clearly seen between anomalous and Anderson localization. In particular, we can observe  much longer tails in the transmission distribution  when  anomalous localization is present. 

{\it Summary and conclusions.-} We have experimentally demonstrated  that  anomalous localization of waves can be induced and controlled in 1D disordered structures. In a microwave waveguide, anomalous localization is produced by the random spacing between dielectric slabs which follow a probability density function with a power-law tail with exponent  $\alpha$, named L\'evy-type distribution. Signatures of  anomalous localization  are recognized by analyzing the random fluctuations of the microwave transmission; thus, we have shown  that the average transmission $\langle T \rangle $ decays with the length $L$ of the waveguides as $L^{-\alpha}$, while $\langle -\ln T \rangle \propto L^\alpha$. In contrast, a stronger wave localization is present in the standard Anderson localization problem that leads to $\langle T \rangle \propto \exp{(-L/\lambda)}$, $\lambda$ being the localization length, and $\langle -\ln T \rangle \propto L$. We point out that anomalous  localization has been experimentally observed in single mode waveguides. For instance, in Ref. \cite{kuhl} an enhancement of localization has been produced by random potentials; however, such anomalous localization is due to an {\em ad hoc} correlated arrangement of scatterers. In our experimental setup, the scatterers are randomly and independently placed inside the waveguide.

Transmission measurements were performed at different microwave frequencies and configurations of the disorder to obtain the complete  transmission distribution. The  experimental results have been well described by our model with the knowledge of two quantities only, the average $\langle -\ln T \rangle$ and the exponent $\alpha$; i.e., all other details of the disorder configuration are  irrelevant for a full statistical description of the transmission. We also note the striking differences between the standard log-normal distribution of the transmission, expected for the case of Anderson localization, and the transmission distributions of our ``anomalous microwave waveguides.''

The understanding of the phenomenon of localization is of fundamental and practical importance. Additionally, because of the universality of this phenomenon in  wave transport in random media, we believe that our work  is of relevance to other research areas, such as quantum electronic transport and  photonics.  Also, the results presented here may be applied to transverse localization \cite{raedt, schwartz,segev}. Moreover, the control of anomalous localization, as shown here, might open new possibilities of taking advantage of the presence of disorder, in the spirit of very recent experimental efforts to manipulate optical properties of random structures \cite{salman, riboli}.

\begin{acknowledgments}

J. C., F. C., J. S.-D., and V. A. G. acknowledge financial support from MINECO (Spain) under the Projects No.  FIS2012-35719-C02-02, No. TEC2010-19751, and No. CSD2008-00066 (CONSOLIDER Program). A.~A. F.-M. and J. A. M.-B. thank Fondo Institucional PIFCA through Grant No. BUAP-CA-169 for partial support.

\end{acknowledgments}


\begin{thebibliography}{50}

\bibitem{anderson} P. W. Anderson, Phys. Rev. {\bf 109}, 1492 (1958).

\bibitem{today} A. Lagendijk, B. van Tiggelen, and D. S. Wiersma, Phys. Today {\bf 62}, 24 (2009), and references therein.

\bibitem{50years} {\it Fifty Years of Anderson localization}, edited by E. Abrahams (World Scientific, Singapore, 2010).

\bibitem{wiersma} D. S. Wiersma, P. Bartoloni, A. Lagendijk, and R. Righini, Nature (London) {\bf 390}, 671 (1997).

\bibitem{stoytchev} M. Stoytchev and A. Z. Genack, Phys. Rev. Lett. {\bf 79}, 309 (1997).

\bibitem{chabanov} A. A. Chabanov, M. Stoytchev,  and A. Z. Genack, Nature (London) {\bf 404}, 850 (2000).

\bibitem{genack} Z. Shi, J. Wang, and A. Z. Genack, Proc. Natl. Acad. Sci. U.S.A. {\bf 111}, 2926 (2014).

\bibitem{penha} A. Pe\~na, A. Girschik, F. Libisch, S. Rotter, and A. A. Chabanov, Nat. Commun. {\bf 5}, 3488 (2014).

\bibitem{yamilov} A. G. Yamilov, R. Sarma, B. Redding, B. Payne, H. Noh, and Hui Cao, Phys. Rev. Lett. {\bf 112}, 023904 (2014).

\bibitem{izrailev} For a review of the topic, see F. M. Izrailev, A. A. Krokhin, and N. M. Makarov, Phys. Rep. {\bf 512}, 125 (2012), and references therein.

\bibitem{elias} I.  Amanatidis, I. Kleftogiannis, F. Falceto, and V.  A. Gopar, Phys. Rev. B {\bf 85}, 235450 (2012). 

\bibitem{soukoulis} C. M. Soukoulis and E. N. Economou, Phys. Rev. B {\bf 24}, 5698 (1981).

\bibitem{evangelou} S. N. Evangelou and D. E. Katsanos, J. Phys. A {\bf 36}, 3237 (2003). 

\bibitem{barthelemy} P. Barthelemy, J. Bertolotti, and D. S. Wiersma, Nature (London) {\bf 453}, 495 (2008). 

\bibitem{salman} S. Karbasi1, R. J. Frazier, K. W. Koch, T. Hawkins, J. Ballato, and A. Mafi, Nat. Commun. {\bf 5}, 3362 (2014).

\bibitem{riboli} F. Riboli, N. Caselli, S. Vignolini, F.  Intonti, K. Vynck, P. Barthelemy, A.  Gerardino, L.  Balet, L.  H. Li, A.  Fiore, M.  Gurioli,  and D. S. Wiersma, Nat. Mater. {\bf 13}, 720 (2014).

\bibitem{falceto_gopar} F. Falceto and V. A. Gopar, Europhys. Lett. { \bf 92}, 57014 (2010).

\bibitem{models} In the incoherent transport regime, similar disorder models have been considered to study statistical properties of the transmission. See, for instance, Refs.~\cite{boose, beenakker, burioni}.

\bibitem{boose} D. Boos\'e and J. M. Luck, J. Phys. A. {\bf 40}, 14045 (2007).

\bibitem{beenakker} C. W. J. Beenakker, C. W. Groth, and A. R. Akhmerov, Phys. Rev. B {\bf 79}, 024204 (2009).

\bibitem{burioni} R. Burioni, L. Caniparoli, and A. Vezzani, Phys. Rev. E {\bf 81}, 060101(R) (2010).


\bibitem{largervalues} We restrict ourselves to $0 < \alpha <1$, where effects of anomalous localization are stronger. However, one can extend the present model to the case $1 < \alpha <2$.

\bibitem{levy} P. L\'evy, {\it Theory of Summation of Random Variables} (Gauthiers-Villars, Paris, 1937).

\bibitem{kolmogorov} B. V. Gnedenko and A. N. Kolmogorov, {\it Limit Distributions for Sums of Independent Random Variables} (Addison-Wesley, Cambridge, MA, 1954).

\bibitem{uchaikin} V. V. Uchaikin and V. M. Zolotarev, {\it Chance and Stability. Stable Distributions and Their Applications} (VSP, Utrecht, 1999), and references therein.


\bibitem{melnikov} V. I. Mel'nikov, Pis'ma Zh. Eksp. Teor. Fiz. {\bf 32}, 244 (1980); [JETP Lett. {\bf 32}, 225 (1980)]. 

\bibitem{abrikosov} A. A. Abrikosov, Solid State Commun. {\bf 37}, 997 (1981).

\bibitem{dorokhov}O. N. Dorokhov, Pis'ma Zh. Eksp. Teor. Fiz. {\bf 36}, 259 (1982); [JETP Lett. {\bf 36}, 318 (1982)].


\bibitem{mello_groups} P. A. Mello, J. Math. Phys. {\bf 27}, 2876 (1986).

\bibitem{markos} P. Marko$\check{\rm s}$ and C. M. Soukoulis, {\it Wave Propagation. From Electrostatics to Photonic Crystals and Left-Handed Materials} (Princeton University Press, Princeton, NJ, 2008).


\bibitem{average} We have experimentally verified that the ensemble average $\langle T \rangle$ is nearly constant within the frequency window of 0.4 GHz around the nominal frequencies 8.5 and 11.5 GHz.


\bibitem{losses} We added  the information of absorption in our  model in a simple way: Assuming a constant absorption,  we rescale the transmissions by the factors 0.8 and 0.75 at the nominal frequencies 8.5 and 11.5 GHz, respectively. We  estimated those factors from transmission measurements. A more realistic model of absorption might improve the  agreement between theory  and experiment.

\bibitem{kuhl} U. Kuhl, F. M. Izrailev, and A. A. Krokhin,  Phys. Rev. Lett. {\bf 100}, 126402 (2008).

\bibitem{raedt} H. De Raedt, A. Lagendijk, and P. de Vries, Phys. Rev. Lett. {\bf 62}, 47 (1989).

\bibitem{schwartz} T. Schwartz, G. Bartal, S. Fishman, and M. Segev, Nature (London) {\bf 446}, 52 (2007).

\bibitem{segev} M. Segev, Y. Silberberg, and D. N. Christodoulides, Nat. Photonics {\bf 7}, 197 (2013).


\end{thebibliography}
\end{document}